\pdfoutput=1
\documentclass[12pt,a4paper]{article}
\usepackage{times}
\usepackage{graphics}
\usepackage{graphicx}
\begin{document}
\date{ }

\begin{center}

{\Large\bf OPTIONS FOR THE NEUTRON LIFETIME MEASUREMENTS IN TRAPS}

\vskip 0.7cm

{\bf Yu. N. Pokotilovski}

\vskip 0.7cm

         {\it Joint Institute for Nuclear Research, Dubna, Russia}

\vskip 0.7cm

 Different geometries for the neutron lifetime measurements by the method of
ultracold neutron storage in material traps and additional possibilities for
the neutron storage in the magnetic storage ring are considered.

\end{center}

\vskip 0.3cm


\vskip 0.2cm

1. INTRODUCTION

 The neutron lifetime is still one of the least accurately measured
fundamental nuclear constants.
 It is known that precision measurement of the neutron lifetime along
with other neutron decay experiments has significant impact on the Standard
Model parameters.
 Recent reviews are published in [1--3].

 There are two principal ways to measure the neutron lifetime.
 In a neutron beam experiment the $\beta$-activity of a beam of cold neutrons is
measured, and the neutron lifetime precision is determined by the accuracy of
absolute measurement of neutron beam density in a well-defined fiducial
volume in the beam and by the accuracy of absolute measurement of the
neutron decay products (protons or electrons) count rate.

 In the ultracold neutron (UCN) storage method the neutron lifetime is
measured by counting surviving neutrons in a magnetic or material trap as a
function of time.
 No absolute measurement is necessary in this case but serious problem arises
in view of UCN losses in traps.
 In the UCN magnetic storage experiments the problem is in possible neutron
spin-flip followed by its escape from the magnetic storage volume.
 Another problem is the marginal neutron trapping due to specular reflection
of superbarrier neutrons from the ideally smooth magnetic mirror leading to
appearance of quasi-bound neutron orbits.

 In material traps the main problem is to account properly for the UCN losses
in their collisions with the surface of a material confinement cavity.
 One tries to minimize these losses choosing the materials with the lowest
neutron capture and suppressing the neutron upscattering by lowering the trap
temperature.

 The value of necessary corrections and systematic errors in inferring the
neutron lifetime from the UCN storage data depend on these losses.

 The present experimental data base for the free neutron decay is large but
contains significant discrepancy between data.
 The Table shows the values of the neutron lifetime
obtained in the beam experiments and in the UCN storage experiments.
 Only the results with uncertainties not exceeding 10 s were taken into
consideration.

\begin{center}
 Results of the neutron lifetime measurements in the beam
experiments and in the UCN storage experiments
\end{center}
\begin{tabbing}
------------------------------------------------------------------------------------------------------------------------\\
Beam experiments qqqqqqqqqqqqqqqqqqqqqqqqqq\=qqqqqqqqqqqqqqqqqqqqq\=\kill
 Beam experiments           \>  Storage experiments \\
------------------------------------------------------------------------------------------------------------------------\\
891$\pm$9 (1988, P. E. Spivak) [4] \>
           \quad 877$\pm$10 (1989, F. Anton et al.) [10] \\
$^{*}$893.6$\pm$3.8$\pm$3.7 (1990, J. Byrne et al.) [5] \>
           \quad 870$\pm$8 (1989, A. G. Kharitonov et al.) [11]\\
889.2$\pm$3.0$\pm$3.8 (1996, J. Byrne et al.) [6]  \>
           \quad 887.6$\pm$3.0 (1989, W. Mampe et al.) [12]\\
$^{*}$886.8$\pm$1.2$\pm$3.2 (2003, M. S. Dewey et al.) [7]\>
           \quad 888.4$\pm$3.3 (1992, V. Nesvizhevsky et al.) [13]\\
$^{*}$886.6$\pm$1.2$\pm$3.2 (2004, J. S. Nico et al.) [8]\>
           \quad 882.6$\pm$2.7 (1993, W. Mampe et al.) [14] \\
886.3$\pm$1.2$\pm$3.2 (2005, M. S. Dewey et al.) [9]\>
           \quad 885.4$\pm$0.9$\pm$0.4 (2000, S. Arzumanov et al.) [15] \\
        \> \quad 881.$\pm$3.0 (2000, A. Pichlmaier et al.) [16] \\
        \> \quad 878.5$\pm$0.7$\pm$0.3 (2004, A. Serebrov et al.) [17] \\
        \> \quad $^{*}$ 874.6$^{+4}_{-1.6}$ (2007, V. Ezhov et al.) [18] \\
        \> \quad 878.2$\pm$1.9 (2009, V.Ezhov et al.) [19] \\
------------------------------------------------------------------------------------------------------------------------\\
Averaged value\>
           \quad Average value without [17] and [19] \\
 887.6$\pm$2.7                    \>
           \quad 885.$\pm$0.82         \\
------------------------------------------------------------------------------------------------------------------------\\
        \> \quad Average value including [17] and [19] \\
        \> \quad 881.3$\pm$0.53                                \\
-------------------------------------------------------------------------------------------------------------------------
\end{tabbing}

 The data marked with the $^{*}$ are not used for obtaining the average
values, only the last results of the corresponding experiment are taken into
consideration.

 The world average value [20] based on [4,6,9,12,13,14,15] is 885.7$\pm$0.8 s.

 It is seen that the most precise UCN storage results may be divided in
two groups: one -- Refs.[12-15] and another -- Refs.[17,19], with the work
[16] between these groups.
 In view of this disagreement between UCN storage measurements and some difference
between the results of the beam and the UCN storage experiments, new precision
measurements of the neutron lifetime by different methods are desirable.
 It is essential that in the UCN storage experiments the total neutron
disappearance probability is measured, whereas the beam experiments are only
sensitive to the neutron beta decay.

\vspace{1mm}

2. DIFFERENT FORMS OF THE UCN MATERIAL TRAPS

\vspace{1mm}

\begin{figure}
\begin{center}
\resizebox{18cm}{13cm}{\includegraphics[width=\columnwidth]{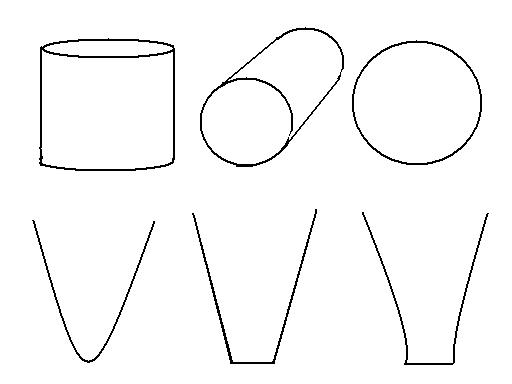}}
\end{center}
\caption{The forms of the UCN traps.}
\end{figure}

 Different geometries (Fig. 1) of storage volume are considered here:
cylindrical with vertically and horizontally directed axis of the cylinder in
respect to the Earth gravitational field, spherical, paraboloidal, conical,
and vase.
 The quasi-spherical and horizontal cylinder geometries have been used in the
neutron lifetime measurements [11,13,17], the vertical cylinder
has been used in the first neutron lifetime experiments with UCN [21],
the conical, vase and paraboloidal forms have never been used and are
considered for comparison.

 In the presence of gravity, for equilibrium distribution of ultracold
neutrons in the available phase space in a cavity, the neutron
loss rate (s$^{-1}$) is determined by the equation (we follow here
the approach of [22]):
\begin{equation}
d\int \varrho({\bf r,v},t)d^{3}{\bf r}\;d^{3}{\bf v}=
-\int\mu({\bf v})({\bf v}{\bf n_{S}})\varrho({\bf r},{\bf v},t)\;
d^{3}{\bf v}\;dS\;dt.
\end{equation}

 Here, $\varrho({\bf r},{\bf v},t)$ is the UCN phase space density,
${\bf n_{S}}$ is the normal to the surface at the point of neutron collision
with the wall.
 The neutron loss probability per one collision $\mu(v)$, for
isotropic neutron incidence, according to the accepted formalism is:
\begin{equation}
\bar\mu(v)=2\,\eta\frac{\kappa(y)}{y^{2}},\quad
\kappa(y)=\arcsin{y}-y\sqrt{1-y^{2}},\quad y=v/v_{b}=\sqrt{E/V},
\end{equation}
where $v$ and $E$ are the neutron velocity and energy, respectively, $v_{b}$
is the neutron boundary velocity of the cavity, $\eta$ is the UCN loss
coefficient determined by the properties of the wall material:
\begin{equation}
\eta=-\mbox{ Im }U/\mbox{ Re }U, \quad U=(\hbar^{2}/2m)4\pi\sum_{i}N_{i}b_{i},
\quad V=\mbox { Re }U,\quad\mbox{ Im }b=-\sigma/2\lambda.
\end{equation}
 Here the complex potential $U$ describes UCN interaction with walls,
$m$ is the neutron mass, $N_{i}$ is the number density of nuclei of type $i$
in the material, $b_{i}$ is the corresponding coherent scattering length on a
bound nucleus of the wall, and $\sigma$ is the cross section of inelastic
processes (capture plus upscattering) for neutrons with wavelength $\lambda$.

 The neutron density depends on the height coordinate $z$ defined relative to
the lowest point of a cavity:
\begin{equation}
\varrho(z,v)=c\; \delta(v_{0}^{2}-2gz-v^{2}),
\end{equation}
where $c$ is the normalization constant, $v_{0}$  and
$\varrho(0,v)=c\;\delta(v_{0}^{2}-v^{2})$ are the neutron velocity and the
neutron density at the bottom of the trap, $g=9.80665$\,m/s$^{2}$ is
gravitational acceleration.

 Solution of Eq. (1) gives the UCN loss rate for different forms of the
traps.

 General expression for the UCN loss probability (s$^{-1}$) of the neutron
with velocity $v$ at the bottom of the cavity is
\begin{equation}
\tau^{-1}(y)=\eta\gamma(y); \qquad \gamma(y)=v_{b}\frac{S(y)}{V(y)},
\end{equation}
where  $S(y)$
is the surface loss term and $V(y)$ is the effective volume.

 In a cylindrical trap with horizontal axis the expressions for $S(y)$ and
$V(y)$ are:
\begin{eqnarray}
S(y)=S_{\rm cyl}(y)+2S_{\rm pl}(y),\quad S_{\rm cyl}(y)=
R\,H\int_{0}^{\varphi_{0}}\kappa
\Biggl(\sqrt{y^{2}-2gR(1-\cos{\varphi})/v_{b}^{2}}
\Biggr)\;d\varphi, \nonumber\\
S_{\rm pl}(y)=\int_{0}^{z_{0}}\kappa\Biggl(\sqrt{y^{2}-2gz/v_{b}^{2}}\Biggr)\;
\sqrt{2Rz-z^{2}}\;dz, \nonumber\\
V(y)=2\,H\int_{0}^{z_{0}}\sqrt{y^{2}-2gz/v_{b}^{2}}\;\sqrt{2Rz-z^{2}}\;dz,
\end{eqnarray}
where $S_{\rm cyl}(y)$ is the loss contribution of the cylinder surface,
$S_{\rm pl}(y)$ is the loss contribution of each of two plane surfaces,
$\varphi_{0}=\arccos\Bigl(1-\frac{(yv_{b})^{2}}{2gR}\Bigr)$,
$z_{0}=(yv_{b})^{2}/2g$, $R$ and $H$ are the radius and the length
of the horizontal cylinder, respectively.

 For a spherical trap of radius $R$ these expressions have the form:
\begin{equation}
S(y)=\pi R\int_{0}^{z_{0}}\kappa\Biggl(\sqrt{y^{2}-2gz/v_{b}^{2}}
\Biggr)\;dz,\quad V(y)=\pi\int_{0}^{z_{0}}\sqrt{y^{2}-2gz/v_{b}^{2}}\;
(2Rz-z^{2})\;dz.
\end{equation}

 For a conical cavity with the angle $\theta$ and flat bottom of radius
$r_{0}$ the loss rate is determined by:
\begin{eqnarray}
S(y)=S_{\rm con}(y)+S_{\rm bot}(y),\quad S_{\rm con}(y)=
\pi\int_{0}^{z_{0}}\kappa\Biggl
(\sqrt{y^{2}-2gz/v_{b}^{2}}\Biggr)\frac{z\;\tan(\theta)}
{\cos{\theta}}\;dz, \nonumber\\
S_{\rm bot}(y)=\frac{\pi\;r_{0}^{2}}{2}\kappa(y),\quad
V(y)=\pi\int_{0}^{z_{0}}\sqrt{y^{2}-2gz/v_{b}^{2}}(z\,\tan{\theta})^{2}\;dz,
\end{eqnarray}
where $S_{\rm con}(y)$ is the conical surface losses and $S_{\rm bot}(y)$
is the bottom surface losses.

 For a vertical cylinder trap with the radius $R$ the loss rate is determined
by:
\begin{eqnarray}
S(y)=S_{\rm cyl}(y)+S_{\rm bot}(y),\quad S_{\rm
 cyl}(y)=\pi\;R\int_{0}^{z_{0}}\kappa\Biggl
(\sqrt{y^{2}-2gz/v_{b}^{2}}\Biggr)\;dz, \nonumber\\
S_{\rm bot}(y)=\frac{\pi\;R^{2}}{2} \kappa(y), \quad
V(y)=\pi\;R^{2}\int_{0}^{z_{0}}\sqrt{y^{2}-2gz/v_{b}^{2}}\;dz,
\end{eqnarray}
where $S_{\rm cyl}(y)$ is the contribution of the cylinder surface.

 For a paraboloidal cavity ($z=r^{2}/R$) the loss rate is determined by
\begin{equation}
S(y)=\pi\int_{0}^{z_{0}}\kappa\Biggl(\sqrt{y^{2}-2gz/v_{b}^{2}}\Biggr)\;
\Biggl(Rz+\frac{R^{2}}{4}\Biggr)\;dz,\quad
V(y)=\pi R\int_{0}^{z_{0}} \sqrt{y^{2}-2gz/v_{b}^{2}}\;z\;dz.
\end{equation}

 For a cavity in the form of a vase ($z=(Rr)^{1/n}-b$) with the flat bottom of
radius $r_{0}$ and radius $r_{m}$ at the height $H$ the surface loss and the
effective volume are:
\begin{eqnarray}
S(y)=S_{\rm vase}(y)+S_{\rm bot}(y),\nonumber\\
S_{\rm vase}(y)=\frac{\pi}{R}\int_{0}^{z_{0}}
\kappa\Biggl(\sqrt{y^{2}-2gz/v_{b}^{2}}\Biggr)(z+b)^{n}\;
\Biggl[1+\frac{n^{2}(z+p)^{2(n-1)}}{R^{2}}\Biggr]\;dz, \nonumber\\
S_{\rm bot}(y)=\frac{\pi r_{0}^{2}}{2}\kappa(y), \nonumber\\
 V(y)=
\frac{\pi}{R^{2}}\int_{0}^{z_{0}}(z+p)^{2n}\sqrt{y^{2}-2gz/v_{b}^{2}}\;dz.
\end{eqnarray}

 Here, $R=H^{n}/(r_{m}^{1/n}-r_{0}^{1/n})$ and $p=(Rr_{0})^{1/n}$.

\begin{figure}
\begin{center}
\resizebox{18cm}{12cm}{\includegraphics[width=\columnwidth]{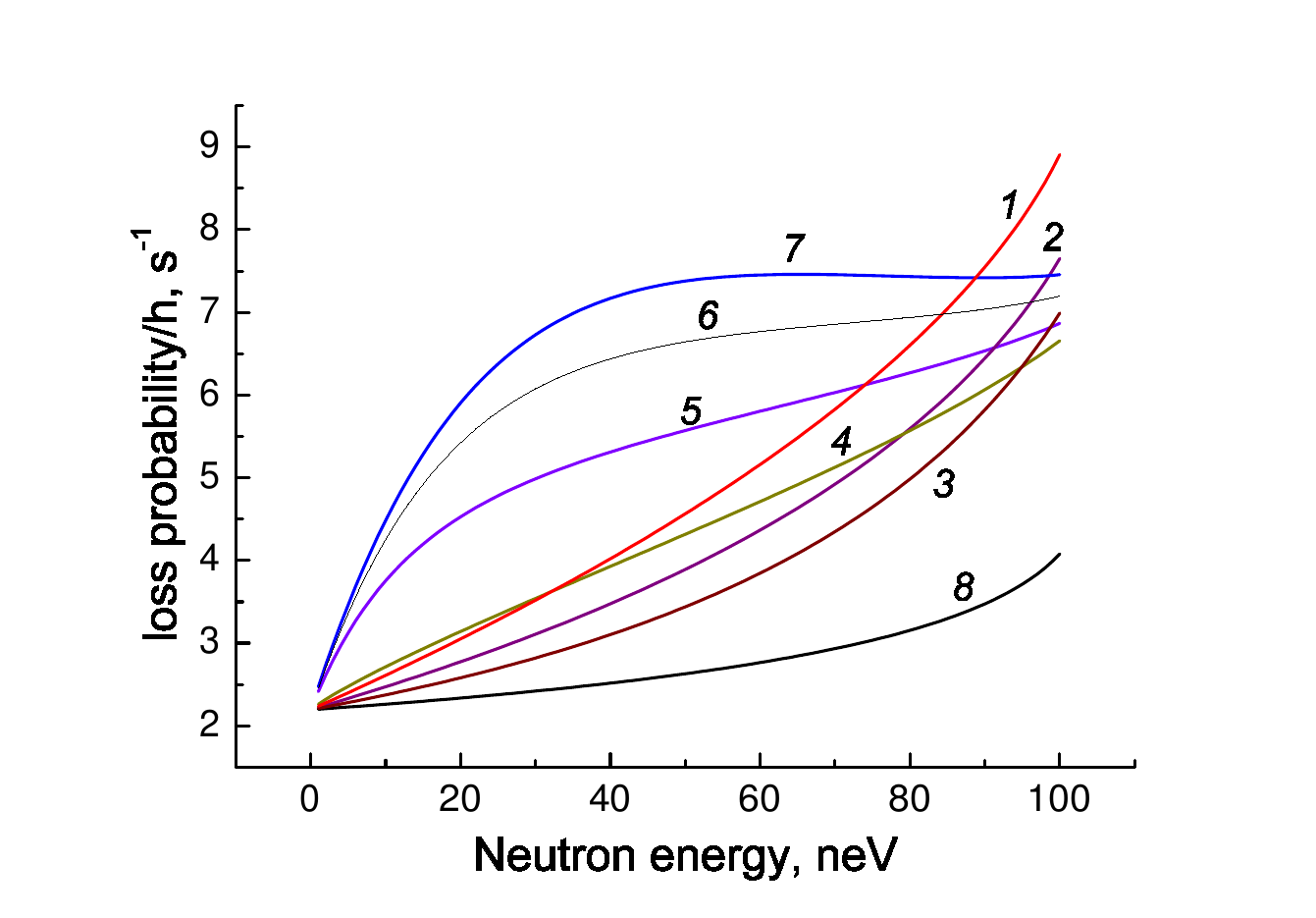}}
\end{center}
\caption{The reduced UCN loss probability (s$^{-1}$) as a function
of neutron energy for different forms of neutron traps:
{\it 1} -- vertical cylinder, R = 50 cm,
{\it 2} -- horizontal cylinder with radius of 50 cm and length 100 cm,
{\it 3} -- sphere, R = 50 cm,
{\it 4} -- paraboloid $z=r^{2}/25$,
{\it 5} -- truncated cone, $\theta=\pi/6$, the bottom radius $r_{0}$ = 5 cm,
{\it 6} -- truncated cone, $\theta=\pi/4$, the bottom radius $r_{0}$ = 5 cm,
{\it 7} -- vase cavity ($z=\sqrt{Rr}-p$), with bottom radius of 5 cm,
top radius of 50 cm and height of 100 cm,
{\it 8} -- infinite plane.}
\end{figure}

 Fig. 2 shows the reduced (the loss coefficient $\eta=1$) UCN loss
probability (s$^{-1}$) as a function of neutron energy at the bottom of
the traps of different form: cylindrical with vertically and horizontally
directed axis of the cylinder in respect to the Earth gravitational field,
spherical, paraboloidal, conical, and vase.
 For comparison the loss probability is shown also for neutrons bouncing at
the infinite horizontal plane.

 All results are shown for $v_{b}$ = 447 cm/s - the boundary velocity for the
fully fluorinated polyether oils (perfluoropolyethers -- PFPE) used in the
recent UCN lifetime measurement and having good neutron reflection properties
[12,16,17].

 As is seen, there is significant difference in the values and energy
dependence of the neutron losses in the traps of similar size but different
form.
 The traps expanding to the top, especially conical ones and in the form of
vase, have low energy dependence of losses in a wide energy range.

\vspace{1mm}

3. EXTRACTION OF THE NEUTRON DECAY CONSTANT

\vspace{1mm}

 The total loss rate $\tau_{\rm st}^{-1}$ of neutrons stored in a trap is
determined by the neutron decay constant $\tau^{-1}_{\rm decay}$ and the wall
loss rate $\tau^{-1}_{\rm loss}$:
\begin{equation}
\tau^{-1}_{\rm st}=\tau^{-1}_{\rm decay}+\tau^{-1}_{\rm loss}.
\end{equation}

 There are two methods of extracting the neutron decay constant from storage
measurements in the presence of losses.

 The first one consists in exclusion of the UCN surface losses from the
experimental $\tau^{-1}_{\rm st}$ using several traps with the same UCN loss
coefficient $\eta$ but with different UCN free path lengths between collisions
with walls:
\begin{equation}
\tau^{-1}_{{\rm st},i}=\tau^{-1}_{\rm decay}+\eta\gamma_{i},
\end{equation}
where $\tau^{-1}_{{\rm st},i}$ and $\gamma_{i}$ are the inverse experimental
storage time and the calculated reduced (without $\eta$) loss in $i$-th trap,
respectively.

 The first method is realized introducing additional surface at the same
storage volume [21] or using the traps of same form but different
size (size extrapolation) [12,14,16,17].

 For extrapolation the measured $\tau^{-1}_{{\rm st},i}$ are plotted against
calculated $\gamma_{i}$ or against calculated inverse free path length
$\lambda^{-1}$ between neutron collisions with walls to obtain the intercept
with the $\gamma=0$ - axis.

 In the second method the UCN storage is performed in the same trap but with
a sequence of different neutron energy spectra -- the energy extrapolation
method [12,13,17].

 Sharp energy dependence of losses is essential for extracting the neutron
decay constant from the storage data by the energy extrapolation method.
 Weak energy dependence of losses on the neutron energy gives (for example
in the energy range from 30 to 100 neV for the curve {\it 7} at Fig. 2)
relative insensitivity of the storage time to the neutron spectrum.
 Conical and vase traps with their weak energy dependence of losses can be
used more efficiently with the size extrapolation.

 One key systematic effect, that may arise when using the size extrapolation
method with two or more neutron traps, is possible difference in the value of
neutron losses at the surface of different traps.
 On the other hand in the energy extrapolation method, applied to the UCN
storage data obtained in the same neutron trap, the homogeneity of the neutron
reflecting coverage of the trap surface is not less important.

 Contrary to the size extrapolation the energy extrapolation needs more
precise information on the realistic energy dependence of
collision neutron losses.
 The energy dependence of UCN losses has been measured twice: first for copper
surface in [23] and then with better precision for the Fomblin oil and
grease surfaces in [24].
 No deviation has been found in both measurements from the energy dependence
of Eq. (2) except for the increase in losses for the neutrons with energy in
the vicinity of the boundary potential of Fomblin oil [24].
 This could be caused by additional small UCN upscattering by liquid surface
modes of Fomblin.

 Actually, it is better to combine both methods of extrapolation in the
neutron lifetime experiment in material traps [17].

\begin{figure}
\begin{center}
\resizebox{18cm}{12cm}{\includegraphics[width=\columnwidth]{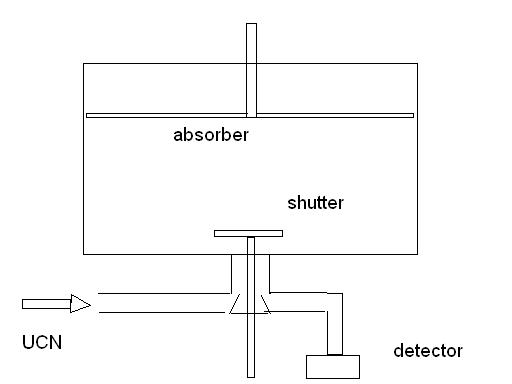}}
\end{center}
\caption{The scheme of the UCN neutron lifetime measurement.}
\end{figure}

 The scheme considered here in more details is shown in Fig. 3.
 The storage measurements are performed in two storage traps of different size
(e.g., two vertical cylinders with the diameters of 50 and 100 cm) for a
sequence of energy spectra of stored neutrons: with the upper energy cut-off
from the lowest possible, say 10 neV, up to the largest possible for the used
material, e.g. $\sim$100 neV for fluoropolymers.
 At small (100--200 s) storage time the initial integral spectra of stored
neutrons is measured, at long storage ($\sim$1000--2000 s) the measured storage
time comprises the neutron decay as well as the energy dependent losses.

\vspace{1mm}

4. THE UCN SPECTRA FORMATION

\vspace{1mm}

 Sharp cutoff in the spectrum may be obtained with the help of a neutron
absorber placed horizontally in the upper part of the storage volume
in the cavity.
 The quality of the UCN absorber is determined by the height of the potential
of the absorbing material.
 Traditionally, polyethylene (CH$_{2})_{n}$ with its comparatively low
negative potential for neutrons $U$(CH$_{2})_{n}\approx -9$\,neV, was considered
as an ``ideal'' absorber.

 Metal absorbers with low or negative scattering length like Ti or
V may be not the best choice as the most effective UCN absorbers
in view of oxidation of its surface.
 Even very thin -- dozens \AA  -- oxide film forming positive potential barrier
entails significant UCN reflection from the surface.
 Fig. 4 shows the neutron energy dependence of reflectivity at isotropic
neutron incidence for polyethylene together with two other possible materials
with good compensation of positive and negative coherent scattering lengths of
nuclei -- glyceride of melyssinic acid C$_{96}$H$_{188}$O$_{6}$ with the height
of the potential $U=-3.3$\,neV and Zirconium hydride ZrH$_{2}$ with almost
perfect compensation of the scattering lengths: $U=-0.325$\,neV.

\begin{figure}
\begin{center}
\resizebox{18cm}{12cm}{\includegraphics[width=\columnwidth]{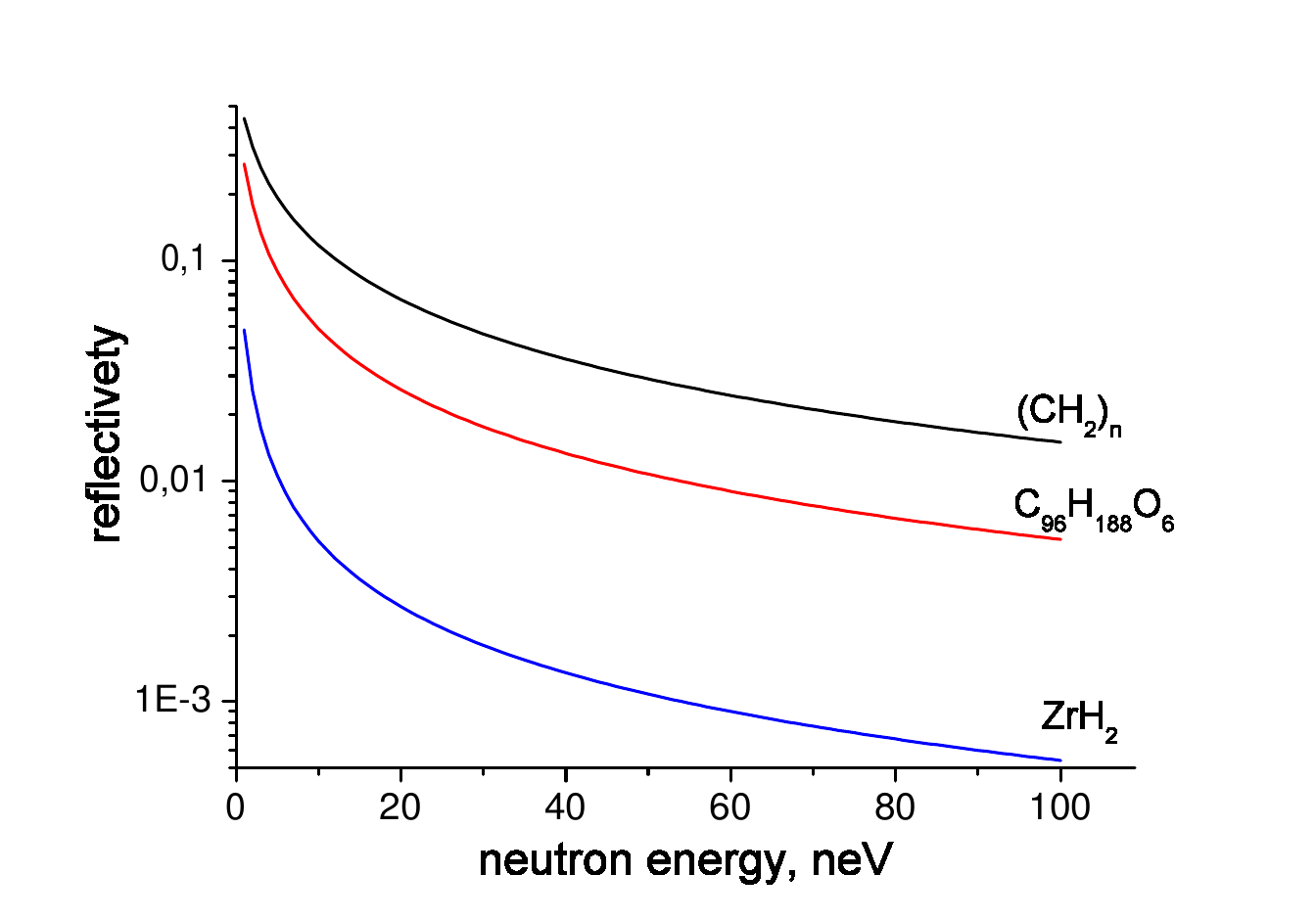}}
\end{center}
\caption{The energy dependence of the UCN reflection probability
from polyethylene (CH$_{2})_{n}$, glyceride of melissynic acid
C$_{96}$H$_{188}$O$_{6}$ and Zirconium hydride ZrH$_{2}$ at isotropic neutron
incidence.}
\end{figure}

 The characteristic "cleaning time" -- the loss time for UCN with the
energies exceeding the cutoff energy determined by the height of the neutron
absorber, may be calculated in the same formalism as was used for calculations
the neutron loss rate in traps (Eqs. (1)--(5)).

 The absorption probability of the neutron with the energy $E$ incident on a
surface with an ideal step function profile at the angle $\theta$ in respect
to the surface normal is
\begin{equation}
T(E,\theta,E_{b})=\frac{4\cos\theta\,\sqrt{\cos^{2}\theta-(E_{b}/E)}}
{(\cos\theta+\sqrt{\cos^{2}\theta-(E_{b}/E)})^{2}},
\end{equation}
where $E_{b}$ is the boundary energy of the absorber.

 The expression for the absorption time constant for a vertical cylinder trap
with the absorber at the height $H$ with respect to the bottom of the
cylinder, has the view:
\begin{equation}
\tau_{\rm clean}(v,H)=\frac{K_{3}}{K_{1}},
\end{equation}
where
\begin{eqnarray}
K_{1}=(v^{2}-2gH)\int_{0}^{\pi/2}
\frac{4 \sin(2\theta)\cos\theta \Bigl(\cos^{2}\theta\mp
v_{a}^{2}/(v^{2}-2gH)\Bigr)^{1/2}}
{\Bigl[\cos\theta+\Bigl(\cos^{2}\theta\mp v_{a}^{2}/
(v^{2}-2gH)\Bigr)^{1/2}\Bigr]^{2}}d\theta, \nonumber\\
K_{3}=\frac{v^{3}-(v^{2}-2gH)^{3/2}}{3g},
\end{eqnarray}
$v_{a}^{2}=2|E_{b}|/m$, and the signs ``-'' and ``+'' correspond to positive
and negative boundary energies of the absorber respectively.
 In the case of positive boundary energy of the absorber the integration over
angle is performed over angles when $\cos^{2}(\theta)>v^{2}_{a}/(v^{2}-2gH)$.
 The Eqs. (16) are valid when $v^{2}>2gH$.

 Fig. 5 shows the energy dependence of the "cleaning time" for three UCN
absorbers: polyethylene, ZrH$_{2}$, and an ideal absorber (100 \%
loss probability for all neutron energies) in the vertical
cylinder with the absorber placed at the height $H=50$\,cm.
 The "cleaning times" for ZrH$_{2}$ and an ideal absorber are practically
the same.

\begin{figure}
\begin{center}
\resizebox{18cm}{12cm}{\includegraphics[width=\columnwidth]{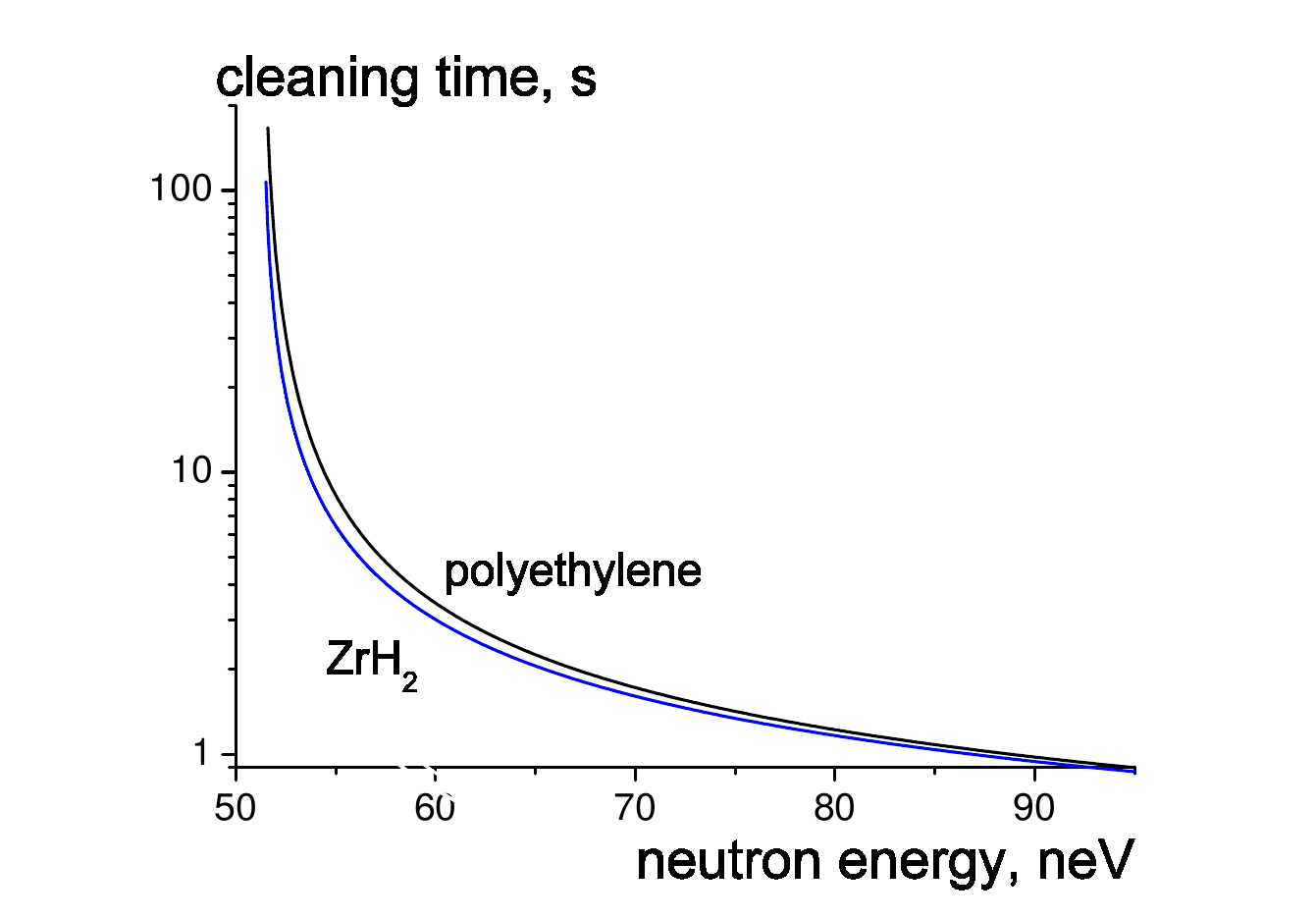}}
\end{center}
\caption{The energy dependence of the UCN cleaning time in a vertical
cylinder with an absorber at H=50 cm for polyethylene,
ZrH$_{2}$ and an ideal absorber.
The latter practically coincides with an ideal absorber.}
\end{figure}

 It is seen that the polyethylene absorber is generally not seriously
inferior in respect to an ideal one except for the energy range close to the
cutoff energy, where the cleaning time for the polyethylene may exceed several
times the cleaning time for an ideal absorber.

 The time needed for cleaning spectrum should not be too large in view of
additional neutron losses during cleaning process.

 In [25] it was proposed to construct a new gravitational trap similar
to described in [17] -- rotating horizontal cylinder with the radius of
120 cm.
 It is obvious that the spectrum cleaning time for such a trap may be
very large.

 The UCN escape from a horizontal cylinder having the longitudinal hole with
the angular width 2$\theta_{0}$ in the upper side of the cylinder may be
imitated by placing the flat vertical ideal absorber as an upward directed
prolongation of the hole.
 The escape rate is calculated according:
\begin{eqnarray}
f=\frac{S(v)}{V_{1}(v)+V_{2}(v)},\quad
S(v)=\frac{2R\cos\theta_{0}+H}{2}
\int_{R(1+\cos\theta_{0})}^{v^{2}/2g}(v^{2}-2gz)\,dz, \nonumber\\
V_{1}(v)=2HR^{2}\int_{0}^{\pi-\theta_{0}}
\sin^{2}\theta\,\sqrt{v^{2}-2gR(1-\cos\theta)}\,d\theta, \nonumber\\
V_{2}(v)=2HR\int_{R(1+\cos\theta_{0})}^{v^{2}/2g}\sqrt{v^{2}-2gz}\;dz,
\end{eqnarray}
where $S(v)$ is the loss term due to collisions with the absorber,
$V_{1}(v)$ and $V_{2}(v)$ are effective volumes of the assumed to
be non-absorbing in this case cylinder and the absorbing
prolongation, respectively.

\begin{figure}
\begin{center}
\resizebox{18cm}{12cm}{\includegraphics[width=\columnwidth]{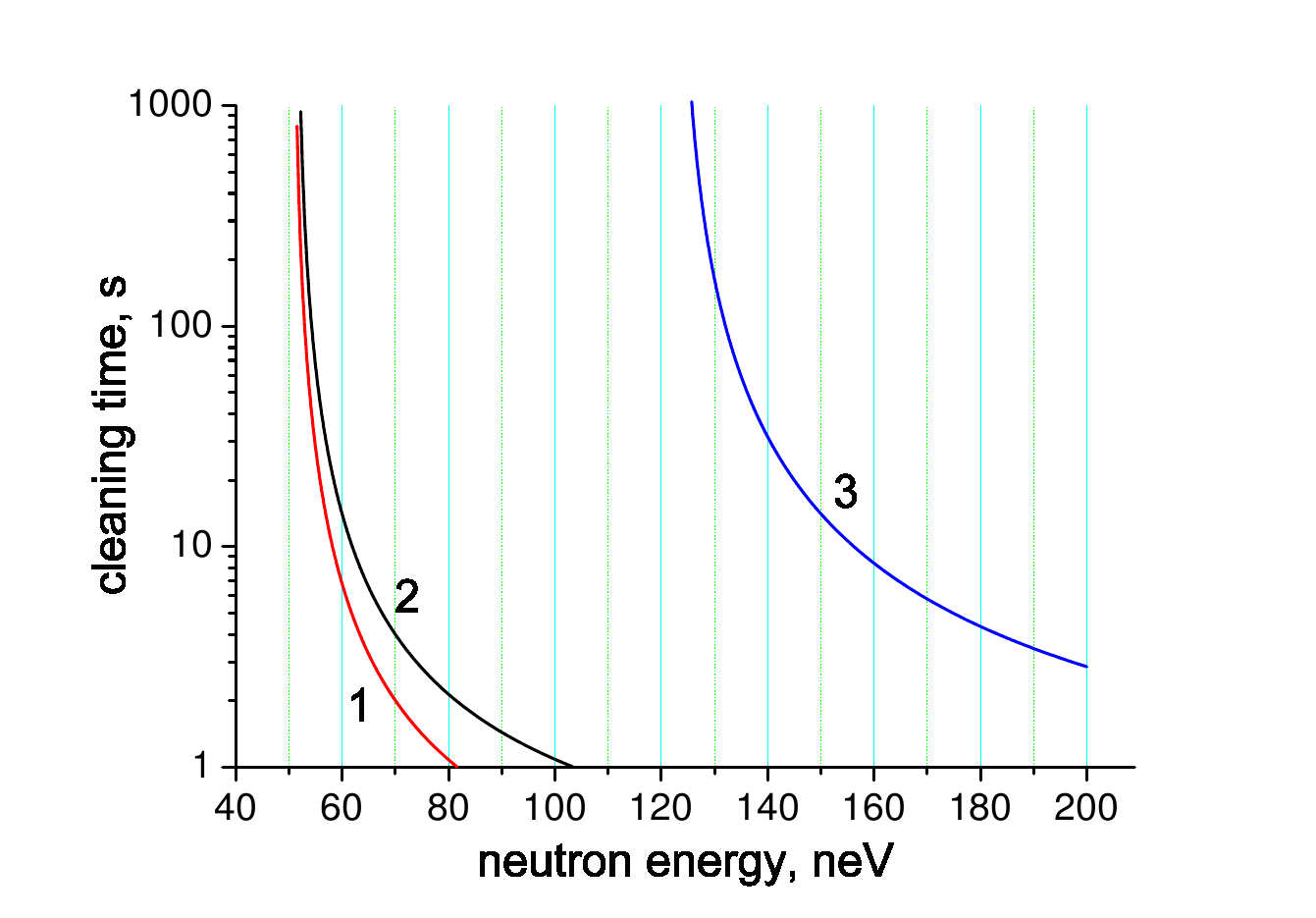}}
\end{center}
\caption{Energy dependence of the calculated UCN escape time from
horizontal cylinders:
1 -- the diameter of 50 cm with the hole in the upper side with
$2\theta_{0}=0.4$ rad,
2 -- the half-cylinder of radius 50 cm,
3 -- the half-cylinder of radius 120 cm.}
\end{figure}

 The neutron escape time from horizontal cylinders is shown in Fig. 6 for
three cases: 2$\theta$=0.4 rad, the cylinder radius 25 cm, and
2$\theta_{0}=\pi$ (open half-cylinder), the cylinder radius 50 cm and
120 cm.

\vspace{1mm}

5. THE SIMULATION OF THE EXPERIMENTS

\vspace{1mm}

 The results of simulation of the experiment with the UCN energy extrapolation
is illustrated in Fig. 7 for the vertical cylinder trap of the radius 50 cm.

\begin{figure}
\begin{center}
\resizebox{18cm}{12cm}{\includegraphics[width=\columnwidth]{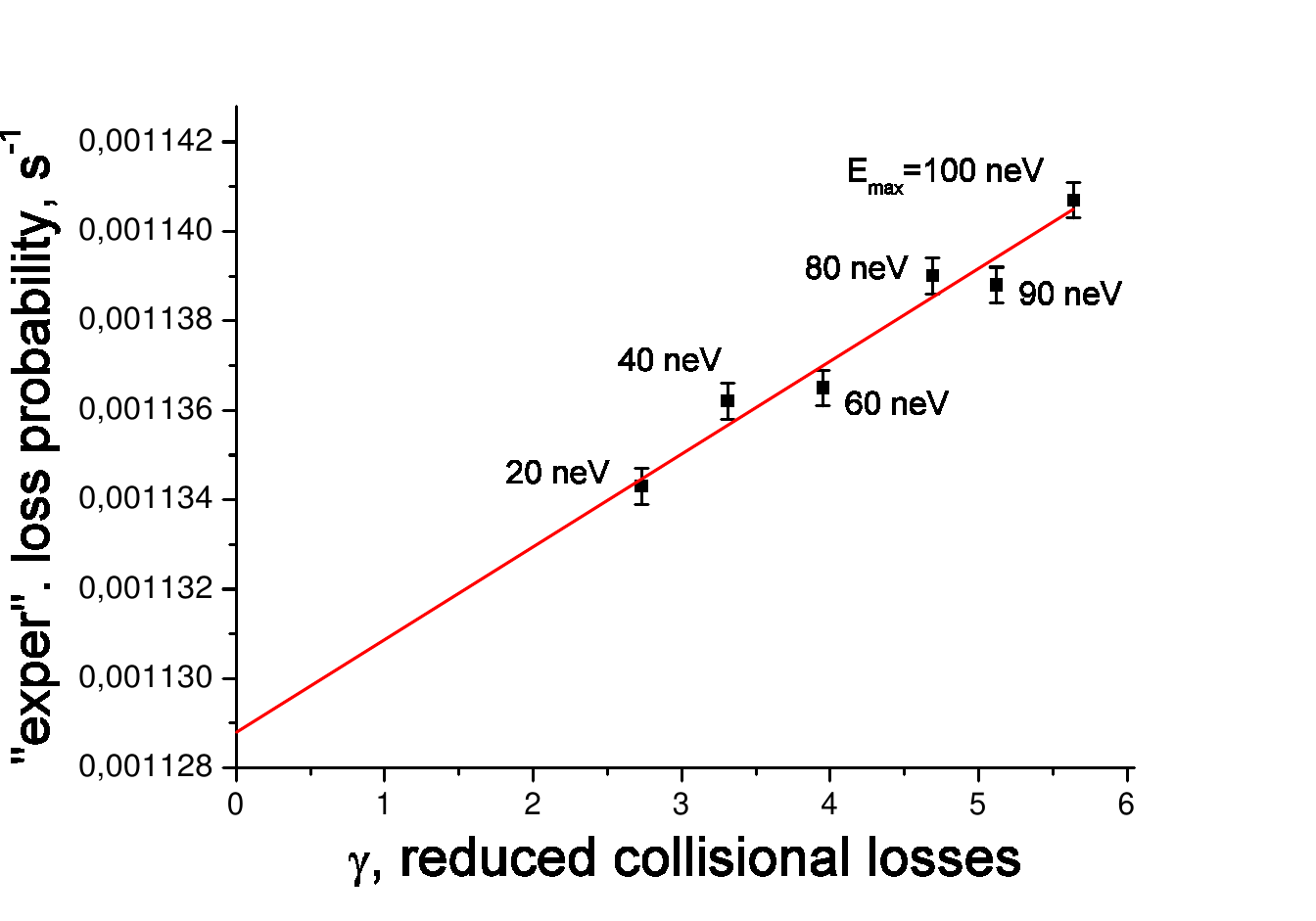}}
\end{center}
\caption{The results of the simulation of the neutron lifetime experiment in
the vertical cylinder trap with radius 50 cm.
 The simulation of the neutron storage was performed for the Maxwellian
spectra with the energy cut-offs at 20, 40, 60, 80, 90 and 100 neV.
 The neutron decay lifetime was 885.7 s.}
\end{figure}

 The storage times and the values of collision losses are calculated for 6
different cut-offs of the initial stored UCN spectrum: from 20 neV to 100 neV.
 The loss coefficient was taken $\eta=2\times 10^{-6}$ [26,17], the
neutron decay lifetime 885.7 s [20], the neutron loss function has the
form of Eq. (2), and it was assumed that initial neutron spectrum at each
measurement is the Maxwell-like tail with perfect cut-off at the corresponding
maximum energy.
 It was assumed that the storage times were measured with a statistical
uncertainty of $4\times 10^{-4}$ (100 and 1000 s measurements, each with
$10^{7}$ counts), and were distributed randomly.
 To reach this statistical precision needs $\sim 530$ cycles of fillings the
neutron storage cavity if the number of stored neutrons in full spectrum with
the $E_{\rm max}$=100 neV is $10^{6}$ per filling and linearly depends on the
cutoff energy.

 In reality extrapolation of storage data to zero losses takes place in
condition of non-complete information about the energy spectrum.
 The incident integral UCN spectrum is measured at small storage time by
changing the height of the absorber.
 But precision of this measurement is not perfect and the incident UCN
spectrum in the trap may differ from the Maxwell-like form.
 Therefore extrapolation to zero losses in an assumption of of the
Maxwell-like spectrum may incur errors.

 Fig. 8 shows the calculated loss rate for different UCN spectra as a function
of the calculated loss rate for the Maxwell-like spectrum.

\begin{figure}
\begin{center}
\resizebox{18cm}{12cm}{\includegraphics[width=\columnwidth]{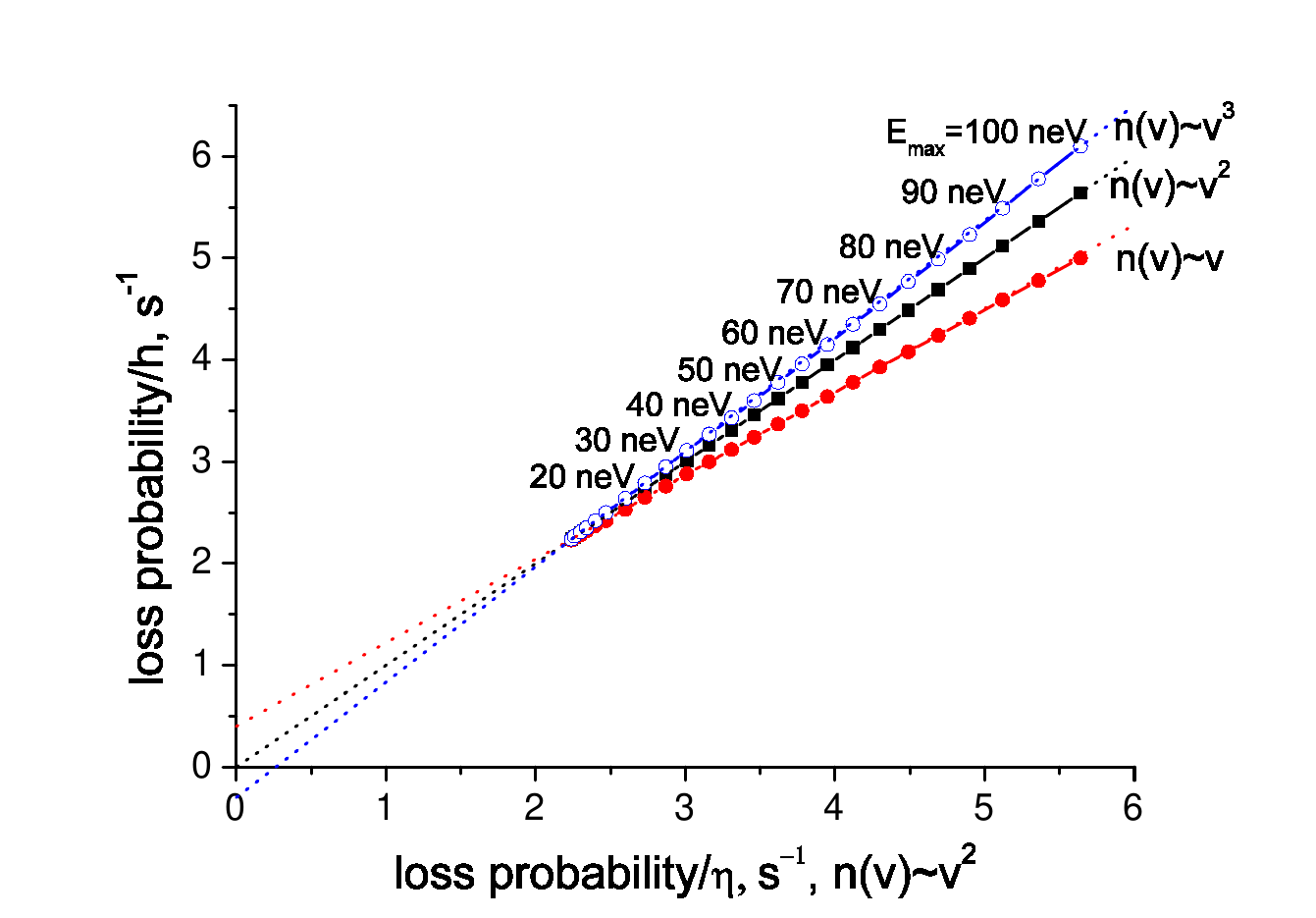}}
\end{center}
\caption{The reduced UCN loss probability in the vertical cylinder trap for
different UCN spectra as a function of reduced UCN loss probability for the
Maxwellian tail spectrum in an assumption of the standard reflection law
of Eq. (2).}
\end{figure}

 The extrapolated losses to zero energy differ from zero in the cases, when
the spectra are not Maxwellian, by the value
$\sim 10^{-6}$\,s$^{-1}\approx 10^{-3}\,\tau_{n}^{-1}$ if the
$\eta=2\times 10^{-6}$.
  It follows from this preliminary estimate that the measurement of the
incident UCN spectra with 5\% precision will decrease this correction to the
level $5\times 10^{-5}\,\tau^{-1}$.

 More serious correction may appear if the real dependence of UCN losses on
the neutron energy differs from the form of Eq. (2) calculated for the ideal
step potential.
 Analysis shows that the UCN loss energy dependence should be determined in
this experiment with the precision no less than 5\%.

 Fig. 9 shows the calculated UCN loss rate for the cases when there is 5\%
admixture to the standard energy dependence of Eq. (2), of different energy
dependent terms in the UCN loss, as a function of the calculated UCN loss for
the standard, Eq. (2) formula.

\begin{figure}
\begin{center}
\resizebox{18cm}{12cm}{\includegraphics[width=\columnwidth]{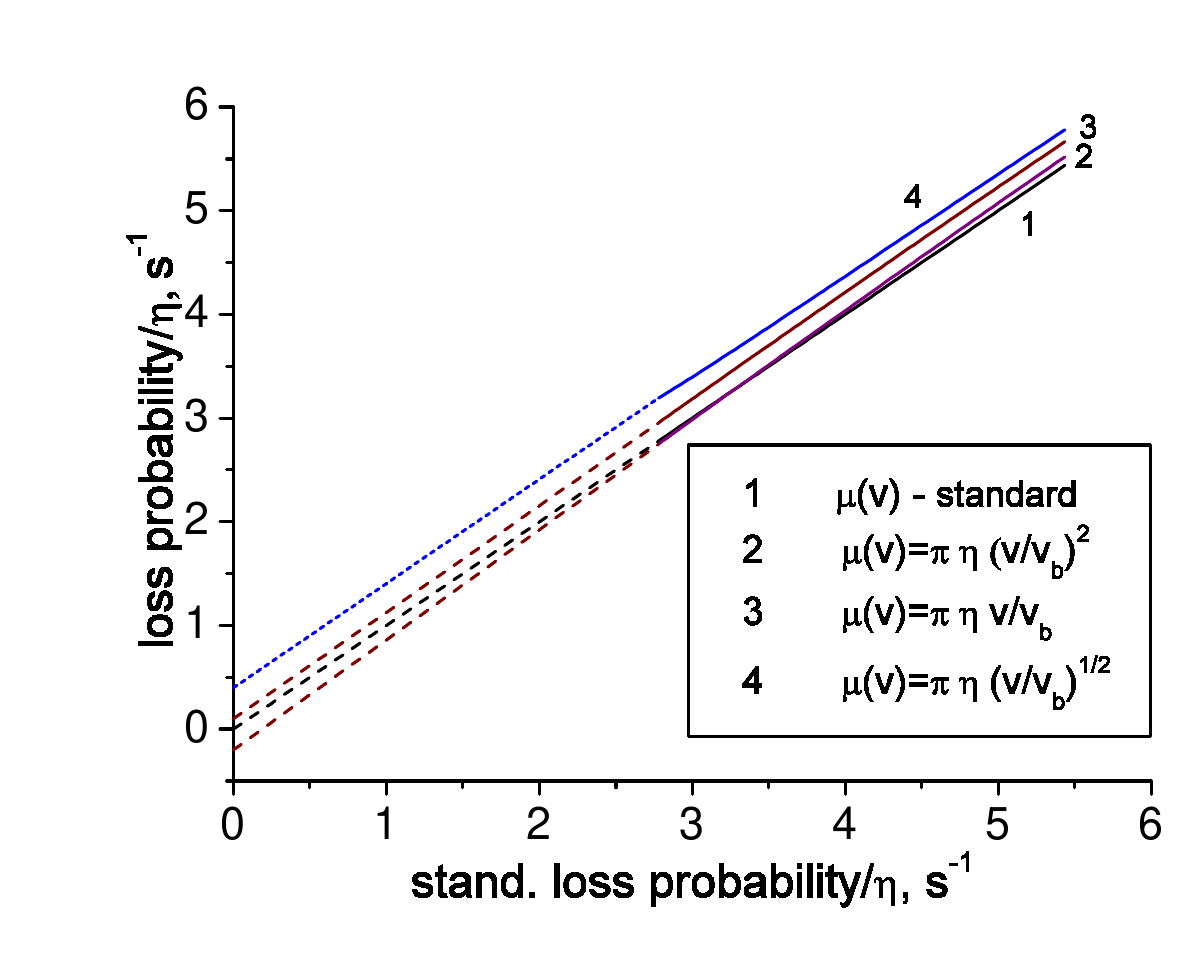}}
\end{center}
\caption{The same but for the Maxwellian tail spectrum and different UCN
dependent UCN losses.}
\end{figure}

 The UCN spectrum was assumed to correspond to the tail of the Maxwell-like
spectrum in these calculations.
 The extrapolated losses to zero energy differ from zero in the cases, when
the neutron losses are not standard, by the value $\sim 10^{-6}
s^{-1}\approx10^{-3}\,\tau_{n}^{-1}$ if the $\eta=2\times 10^{-6}$
in the worst case of the square root UCN energy dependence of the
UCN loss coefficient.

 It is obvious that for higher precision the loss corrections have to be
evaluated more precisely and should be made as low as possible.

 The low loss probability per one neutron collision with the wall is expected
from the result of recent experiment [17], where it was obtained
$\eta\approx 2\times 10^{-6}$ for the low temperature fluorinated
oil (PFPE), and there is hope to decrease further this value below
$10^{-6}$ [27,28].
  On the other hand, larger size storage traps compared to [17] should
decrease the loss rate due to a decrease of the wall collision rate.

6. THE "ACCORDION" TRAP

 The effects of gravity on the storage of UCN in material traps with neutron
losses have been considered first in [22] and in [29] with implications for
experiments to determine the neutron lifetime.
  The consideration in [29] was focused on rectangular traps similar to the
used in the experiments [12,16].
 In these experiments the extrapolation to zero losses was performed by
changing the length of the trap in horizontal direction.

 An interesting proposal in the paper [29] was to use the UCN trap in
the form of a bellows with horizontal axis.
 When the length of bellows and consequently the volume of the trap is
changed, the surface is remained unchanged.
 In this case extrapolation to zero losses is straightforward.
 The project to realize this approach is published in [30].

 The model calculations of the UCN losses in wall collisions as a function of
the length of the bellows are presented below.

 In the absence of gravity the accordion geometry is ideal in the sense of
extrapolations to the infinite volume.
 In the presence of gravity the upper and lower parts of the trap are not
equivalent, and the change of the form of the trap may, in principle, give
some nonzero systematic effect for the value of extrapolated UCN losses.

 The exact geometric form of the surface for different stretches of the
bellows is not known.
 Here two models are used: the linear model (truncated cones) and the
sinusoidal one.
 One segment for the linear model of the bellows is shown in Fig. 10.

\begin{figure}
\begin{center}
\resizebox{18cm}{12cm}{\includegraphics[width=\columnwidth]{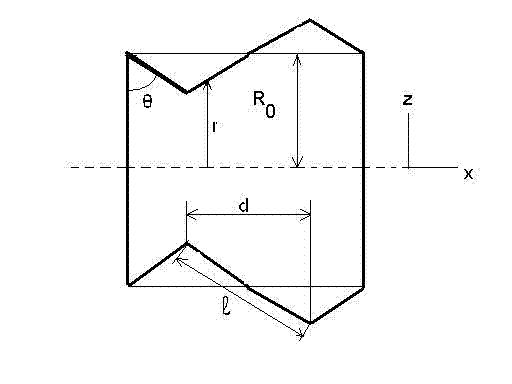}}
\end{center}
\caption{Schematic view of the "accordion" segment.}
\end{figure}

 It is assumed that at the stretching of the bellows the medium radius of the
bellows is unchanged but the outer $R_{\rm max}$ and the inner $R_{\rm min}$
radii are changed according to the following expressions:
\begin{eqnarray}
d=H/N,\quad \sin\theta=d/l,\quad
r=R_{0}+\frac{d}{\tan\theta}\Bigl(\frac{x}{d}-\frac{1}{2}\Bigr), \nonumber\\
 z=r\cos\varphi,\quad
R_{max,min}=R_{0}\pm\frac{d}{2\tan\theta},
\end{eqnarray}
where $H$ is the length of the bellows, $N$ is the number of truncated conic
segments of the bellows, $d$ is the length of the segment in horizontal
direction, $R_{0}$ is the medium radius of the bellows, $l$ is the length of
the generatrix of the conic segment, $\theta$ is the deformation angle,
$\varphi$ is the angle around horizontal axis of the bellows.
 In this case the datum in $z$ direction is at the axis of the bellows.

 Similar to the calculation for other UCN trap forms we have:
\begin{eqnarray}
S(y)=N\cdot S_{\rm con}(y)+2\cdot S_{\rm pl}(y),\nonumber\\
S_{\rm con}(y)=\frac{1}{\sin\theta}\int_{0}^{d}dx
\int_{0}^{\pi}r\kappa\Biggl(\sqrt{y^{2}-2gr\cos{\varphi}/v_{b}^{2}}
\Biggr)\;d\varphi, \nonumber\\
S_{\rm pl}(y)=\int_{-R_{0}}^{R_{0}}\kappa\Biggl(\sqrt{y^{2}-2gz/v_{b}^{2}}
\Biggr)\;\sqrt{2R_{0}z-z^{2}}\;dz, \nonumber\\
V(y)=2\,N\int_{0}^{d}\,dx
\int_{-r}^{r}\sqrt{y^{2}-2gz/v_{b}^{2}}\;\sqrt{2rz-z^{2}}\;dz,
\end{eqnarray}
where $S_{\rm con}(y)$ is the surface loss on one conic section of the bellows,
$S_{\rm pl}(y)$ is the surface losses at the two disc surfaces of the bellows,
$V(y)$ is the effective volume term.

 The results of calculations of the loss rate as a function
of the inverse volume of the bellows is shown in Fig. 11.
 The parameters of the trap were taken from the Ref. [30]:
$l=2.9$\,cm, $R_{0}=(R_{max}+R_{min})/2=26.75$\,cm, the length was varied
between 30 and 120 cm.

\begin{figure}
\begin{center}
\resizebox{18cm}{12cm}{\includegraphics[width=\columnwidth]{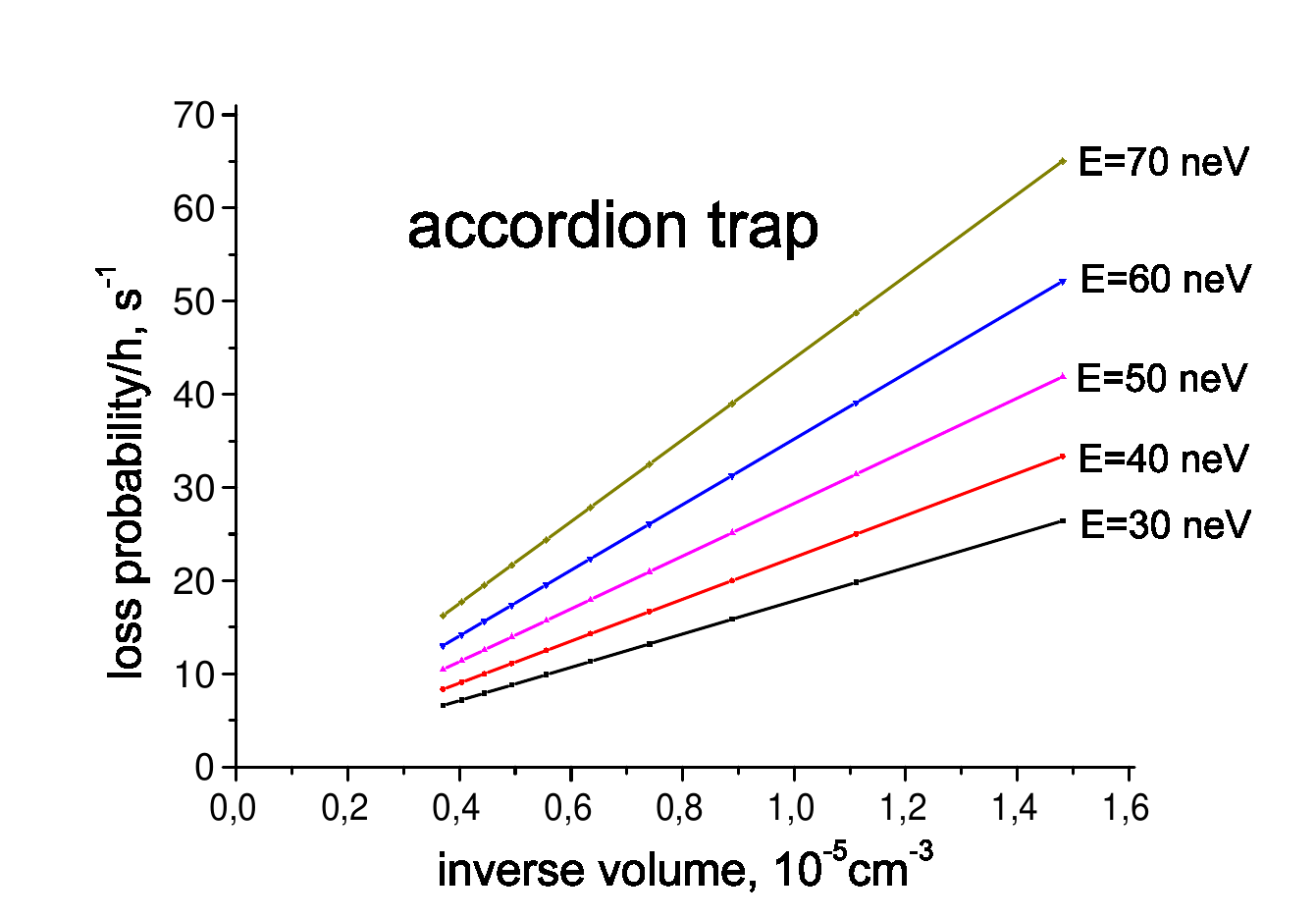}}
\end{center}
\caption{Reduced UCN loss probability in the "accordion" trap for
different UCN energies as a function of the inverse volume of the
trap. The datum is placed at the axis of the trap.}
\end{figure}

 Linear extrapolation $\gamma(1/V)=\gamma_{\inf}+b\times (1/V)$ for each of
the energies gave $\gamma_{\inf}\approx (-5\pm 5)\times 10^{-3}$.
 At the loss coefficient $\eta=10^{-5}$ the systematic effect arising from
the deformation of the UCN storage volume at the size extrapolation is about
$5\times 10^{-8}$ s$^{-1}$ for the inverse lifetime or $\approx 0.04 $s for
the neutron lifetime and does not depend on energy.

 At larger value of $l$ the systematic effect at the extrapolation to infinite
volume is increased.

 The serious problem in the use of bellows UCN trap is deposition of the low
temperature coverage at the corrugated surface and its stability at
the deformation of the bellows.
 Another disadvantage consists in the narrow energy range of confined
neutrons $\approx 30-70$ neV for the datum placed at the axis of the trap
and in the case of the PFPE coverage for the bellows size of [30].
 These limits follow from the conditions that the UCN should have the energy
sufficient to reach the highest points of the trap and not exceed the boundary
energy of the PFPE $E_{b}\approx 106$ neV.

\vspace{1mm}

7. MAGNETIC TRAPS

\vspace{1mm}

 One of the most interesting methods of slow neutron storage has been proposed
and experimentally demonstrated by Paul and collaborators [31,10].
 It was a toroidal hexapole magnetic trap for very slow neutrons in the
$\mu$eV energy range: the neutrons with longitudinal velocities
between 8 and 15 m/s and radial velocity spread up to 4 m/s were
stored in the ring with the diameter of $\sim$ 110 cm.
 The neutron injection from the neutron guide into the magnetic ring was
performed with the help of fast pneumatically driven neutron mirror reflector.
 The final experiments, in which the neutron lifetime was measured with the
precision 10 s, were performed at the very cold neutron channel of the PF-2
neutron source [32] of the ILL High Flux Reactor.

 The great advantage may be achieved using pulsed injection of neutrons from
the pulse neutron source [33,34].
 In this method the neutron bunch from the UCN converter after spreading over
the neutron guide can be trapped in the UCN cavity with high efficiency.

 The detailed MCNP calculations for a solid para-hydrogen converter with an
optimal cold ortho-hydrogen premoderator located in the radial channel of the
MARK-2 TRIGA reactor yield the production of $\sim 10^{8}$ neutrons/Mole/10 MJ
reactor pulse in the neutron velocity half-cylinder with the radius of 4 m/s
and the longitudinal velocity interval from 8 to 15 m/s .
 Possible size of one-mole para-hydrogen converter: the diameter of 7 cm, the
thickness of 0.5 cm is optimal for the reactor radial channel and for the
neutron extraction in this energy range.

 At the distance $L$ from the neutron source to the entrance to the storage
ring, and the neutron velocity interval of stored neutrons between
$v_{1}$ and $v_{2}$, the length of the neutron bunch passed this
distance is $L((v_{2}/v_{1})-1)$.
 The length of the neutron guide at the TRIGA reactor from the moderator to
the entrance of the magnetic ring may be as short as about 5-6 m/s, and at
the length of the storage ring $\sim$350 cm [31] practically optimal
condition may be realized for the neutron injection: almost all the bunch in
the longitudinal velocity range 8-15 m/s fills the storage ring during one
reactor pulse.
 With larger storage ring and consequently larger orbit time the injection and
storage will be more effective.

 The neutron injection efficiency is not reported in [31].
 In an assumption that the neutron survival during transport through the
neutron guide and injection is only 1\% it is expected up to
10$^{6}$ neutrons trapped in the ring per one filling.
 It is three orders of the magnitude larger than reported in [31].

 The additional advantage consists in the possibility to variate the spectrum
of injected neutrons with help of the pulse injector due to time-of-flight
separation of neutrons in the neutron guide.

 Similar possibility of pulsed neutron filling from pulse neutron source
exists for other types of the magnetic traps for ultracold neutrons
[35] if one provides the magnetic trap with pulse magnetic shutter.

 It seems possible, in principle, to control neutron losses from magnetic
ring by placing very slow neutron detector based on the low
temperature scintillator (for example pure CsI [36]) at
the outer side of the ring storage volume and PMT in the central
part of the ring.

 Another possibility -- the decay proton counting is possible placing proton
counters in the center of the storage ring and introducing accelerating
electric field between the walls of the storage volume and the proton
detectors.

The author is grateful to Dr. V. K. Ignatovich for clarifications
concerning his work [22].

\vspace{1mm}

 References

1. H. Abele, Prog. Part. Nucl. Phys. {\bf 60}, 1 (2008).

2. N. Severijns, M. Beck and O. Naviliat-Cuncic, Rev. Mod. Phys.
{\bf 78}, 991 (2006).

3. J. S. Nico and W. M. Snow, Ann. Rev. Nucl. Part. Sci. {\bf 55}, 27 (2005);
nucl-ex/0612022.

4. L. N. Bondarenko, V. V. Kurguzov, Yu. A. Prokof'ev, {\it et al.},
Pis'ma v ZhETF {\bf 28}, 329 (1978) [JETP Lett. {\bf 28}, 303 (1978)];
P. E. Spivak, ZhETF {\bf 94}, 1 (1988) [JETP {\bf 67}, 1735 (1988)].

5. J. Byrne, P. G. Dawber, J. A. Spain, {\it et al.}, Phys. Rev. Lett.
{\bf 65}, 289 (1990).

6. J. Byrne, P. G. Dawber, C. G. Habeck, {\it et al.}, Europhys. Lett.
{\bf 33}, 187 (1996).

7. M. S. Dewey, D. M. Gillian, J. S. Nico, {\it et al.}, Phys. Rev. Lett.
{\bf 91}, 152302 (2003).

8. J. S. Nico, M. S. Dewey, D. M. Gillian, {\it et al.}, nucl-ex/0411041.

9. M. S. Dewey, D. M. Gillian, J. S. Nico, {\it et al.},
Phys. Rev. C {\bf 71}, 055502 (2005).

10. F. Anton, W. Paul, W. Mampe, {\it et al.}, Z. Phys. C {\bf 45}, 25 (1989).

11. A. G. Kharitonov, V. V. Nesvizhevsky, A. P. Serebrov, {\it et al.},
Nucl. Instrum. Methods A {\bf 284}, 98 (1989).

12. W. Mampe, P. Ageron, J. C. Bates, {\it et al.}, Nucl. Instrum. Methods
 A {\bf 284}, 111 (1989); Phys. Rev.  Lett. {\bf 63}, 593 (1989).

13. V. V. Nesvizhevsky, A. P. Serebrov, R. R. Tal'daev, {\it et al.},
ZhETF {\bf 102}, 740 (1992); [JETP {\bf 75}, 405 (1992)].

14. W. Mampe, L. N. Bondarenko, V. I. Morozov, {\it et al.}, Pis'ma v ZhETF
{\bf 57}, 77 (1993); [JETP Lett. {\bf 57}, 82 (1993)].

15. S. Arzumanov, L. Bondarenko, S. Chernyavsky, {\it et al.},
Phys. Lett. B {\bf 483}, 15 (2000).

16. A. Pichlmaier, J. Butterworth, P. Geltenbort, {\it et al.},
Nucl. Instrum. Methods A {\bf 440}, 517 (2000);
A. Pichlmaier, {\it Dissertation} (TU M\"unchen, 1999).

17. A. Serebrov, V. Varlamov, A. Kharitonov, {\it et al.},
Phys. Lett. B {\bf 605}, 72 (2005); nucl-ex/0408009;
 A. P. Serebrov, V. E. Varlamov, A. G. Kharitonov, {\it et al.}
Phys. Rev. C {\bf 78}, 035505 (2008).

18. V. Ezhov, in {\it The Proceeedings of the VI International Conference
``Ultracold and Cold Neutrons, Physics and Sources''}
(St. Petersburg, Moscow, 2007), http://cns.spb.ru/6UCN/proceed.html.

19. V. Ezhov, in {\it The Proceeedings of the VII International Conference
``Ultracold and Cold Neutrons, Physics and Sources''} (St. Petersburg, 2009),
http://cns.spb.ru/7UCN/proceed.html.

20. Rev. Part. Phys., Phys. Lett. B {\bf 667}, 1070 (2008).

21. Yu. Yu. Kosvintsev, Yu. A. Kushnir, V. I. Morozov, {\it et al.}, Pis'ma v
ZhETF {\bf 31}, 257 (1980) [JETP Lett. {\bf 31}, 236 (1980)].

Yu. Yu. Kosvintsev, V. I. Morozov, G. I. Terekhov, Pis'ma v
ZhETF {\bf 36}, 346 (1982) [JETP Lett. {\bf 36}, 424 (1982)].

Yu. Yu. Kosvintsev, V. I. Morozov, G. I. Terekhov, Pis'ma v
ZhETF {\bf 44}, 444 (1986) [JETP Lett. {\bf 44}, 571 (1986)].

22. V. K. Ignatovich, G. I. Terekhov, JINR Commun. P4-9567 (Dubna, 1976).

23. L. V. Groshev, V. N. Dvoretsky, A. M. Demidov, {\it et al.},
JINR Preprint P3-9534 (Dubna, 1976).

24. D. J. Richardson, J. M. Pendlebury, P. Iaydjiev, {\it et al.},
Nucl. Instrum. Methods A {\bf 308}, 568 (1991).

25. A. Serebrov, A. Fomin, I. Shoka, {\it et al.}, ``New project
of neutron lifetime measurement with gravitational trap of UCN.'',
in {\it The Proceedings of the VI-th International Conference
``UltraCold and Cold Neutrons, Physics and Sources''}, (St.
Petersburg, Moscow, 2007), http://cns.spb.ru/6UCN/proceed.html.

26. Yu. N. Pokotilovski, ZhETF {\bf 123}, 203 (2003)
[JETP {\bf 96} 172 (2003)].

27. Yu. N. Pokotilovski, Nucl. Instrum. Methods A {\bf 554}, 356 (2005).

28. Yu. N. Pokotilovski, I. Natkaniec and K. Holderna-Natkaniec,
Physica B {\bf 403}, 1942 (2008).

29. J. M. Pendlebury and D. J. Richardson, Nucl. Instrum. Methods A {\bf 337},
504 (1994).

30. B. Yerozolimsky, A. Steyerl, O. Kwon {\it et al.},
J. Res. Natl. Inst. Stand. Technol. {\bf 110}, 351 (2005); and in {\it The
Proc. XII Intern. Seminar on Interaction of Neutrons with Nuclei}
(Dubna, May 26--29, 2004), p.222.

31. K.-J. K\"ugler, W. Paul, and U. Trinks, Phys. Lett. {\bf 72B}, 422 (1978).

K.-J. K\"ugler, K. Moritz, W. Paul {\it et al.}, Nucl. Instrum. Methods
{\bf 228}, 240 (1978).

32. A. Steyerl, H. Nagel, F.-X. Schreiber, {\it et al.},
Phys. Lett. A {\bf 116}, 347 (1986).

33. Yu. N. Pokotilovski, Nucl. Instrum. Methods A {\bf 356}, 412 (1995).

34 A. Frei, Y. Sobolev, I. Altarev, {\it et al.},
Eur. Phys. Journ. A {\bf 34}, 119 (2007).

35. V. V. Vladimirskii, ZhETF {\bf 39}, 1062 (1960) [Sov. Phys. JETP {\bf 12},
740 (1960)].

Yu. G. Abov, S. P. Borovlev, V. V. Vasil'ev, {\it et al.},
Yad. Fiz. {\bf 38}, 122 (1983) [Sov. J. Nucl. Phys. {\bf 38}, 70 (1983)].

V. F. Ezhov, A. Z. Andreev, A. A. Glushkov {\it et al.},
Journ. Res. NIST {\bf 110}, 345 (2005).

R. Picker, I. Altarev, J. Br\"oker, {\it et al.},
J. Res. NIST {\bf 110}, 357 (2005).

J. D. Bowman and S. I. Pentila, J. Res. NIST {\bf 110}, 361 (2005).

36. J. B. Birks, {\it The Theory and Practice of Scintillation
Counting} (Pergamon Press, Oxford, 1964), p. 460.

\end{document}